\documentclass[pre,showpacs,twocolumn,epsf]{revtex4}
\usepackage{graphicx} 
\usepackage{dcolumn}
\usepackage{pstricks}
\input epsf 

\def\openone{\leavevmode\hbox{\small1\kern-3.3pt\normalsize1}}
\newcommand{\la}{\langle}
\newcommand{\ra}{\rangle}
\newcommand{\be}{\begin{equation}}
\newcommand{\ee}{\end{equation}}
\newcommand{\bea}{\begin{eqnarray}}
\newcommand{\eea}{\end{eqnarray}}
\newcommand{\pd}{\partial}

\begin{document}

\title{Analytical results for a Fokker--Planck equation in the  small noise limit}
\author{Eric Lutz}
\affiliation{Department of Quantum Physics, University of Ulm,
D-89069 Ulm, Germany}
 
\date{\today}

\begin{abstract}
We  present analytical results for the lowest cumulants of a stochastic process described by a Fokker--Planck equation with nonlinear drift. We show that, in the limit of small fluctuations, the mean,  the variance and the covariance of the process can be expressed in  compact form with the help of the Lambert $W$ function. As an application, we  discuss the interplay of noise and nonlinearity far from equilibrium. 
\end{abstract}
  
\maketitle
\section{Introduction}
The Fokker--Planck equation is a basic equation in the theory of stochastic processes. Since its introduction by Adriaan Fokker (1914) and Max Planck (1917) at the beginning of the last century, it has become  a very useful tool for characterizing the  time evolution of systems subjected to fluctuations and noise \cite{ris89}.  The Fokker--Planck equation is a partial differential equation of first order in time for the probability density function in position $P(x,t)$ of the system. By definition, the quantity $P(x,t)\,dx$ is the probability to find the position of the system in the interval $[x,x+dx]$ at time $t$. In one dimension, the Fokker--Planck equation takes the form
 \be
 \label{eq1}
 \frac{\pd P(x,t)}{\pd t} =
 -\frac{\pd}{\pd x}\left[K(x)P(x,t)\right] + D \frac{\pd^2 P(x,t)}{\pd
 x^2} \ ,
 \ee
where $K(x) $ is the drift and $D$ the diffusion coefficient \cite{rem}. These two quantities have a simple physical interpretation. The drift coefficient, $K(x)=-U'(x)$, corresponds to the force experienced by the system moving in the external potential $U(x)$. It describes the deterministic component of the dynamics, that is, the evolution in the absence of fluctuations. On the other hand, the diffusion constant $D$ measures the intensity of the noise and represents the stochastic part of the diffusion process.

The Fokker--Planck equation (\ref{eq1}) is not solvable analytically for arbitrary nonlinear drift coefficients $K(x)$. Closed--form solutions for the probability distribution $P(x,t)$ are only known for very simple forms of the drift term \cite{ris89}. One example is the Wiener process which describes  a free particle in the presence of noise. Here the drift coefficient vanishes and Eq.~(\ref{eq1}) reduces to the  simple diffusion equation. Another exactly solvable example is  a noisy harmonically bound particle  which corresponds to a linear drift, $K(x)= -a x$. This random  process is known as the Ornstein--Uhlenbeck  process.  In the general case, however, approximate solution methods are essential for practical applications. Among these,  the small noise approximation is an important approximation scheme which can be applied when the fluctuations are weak. In this case, the dynamics of the system is mostly governed by the drift term and the noise can be treated as a small perturbation. As a result, one can obtain approximate expressions for the lowest cumulants of the probability density function, such as the mean and the variance of the stochastic process. However, in order to be usefully implemented, the small noise expansion requires the analytical solution of the deterministic equation, $\dot x(t) = K(x)$, in the absence of noise. Unfortunately, an exact solution of this equation is usually not available. 

The purpose of this paper is to present a simple, yet non trivial, example where the small noise approximation can be implemented analytically. Specifically, we shall consider  Eq.~(\ref{eq1}) with the nonlinear drift coefficient
\be
\label{eq01}
 K(x)= - \frac{x}{1+x^2}\ .
\ee
Fokker--Planck equations of this type appear in the semiclassical description of a very efficient laser cooling method known as Sisyphus cooling \cite{coh98,cas91}. The 1997 Nobel prize was awarded to Steven Chu, Claude Cohen-Tannoudji and William  Phillips for the development of these new laser cooling methods. In addition, these ordinary  Fokker--Planck equations   play an important role in the investigation of unusual statistical mechanical properties of cold atoms, such as ergodicity breaking \cite{lut04}. As we shall see, the solution of the deterministic equation of motion with the drift coefficient (\ref{eq01}) can be expressed explicitely in terms of the Lambert $W$ function.  The Lambert function, named after Johann Heinrich Lambert (1728--1777),  is the solution of the transcendental equation $W(x) \exp[W(x)] =x$ \cite{cor96}. The function $W(x)$ can thus be viewed as a simple generalization of the logarithm. It has recently been realized that many physical problems can  be solved analytically using  the Lambert function, making their analysis more tractable. Examples include problems from electrostatics and quantum mechanics  \cite{val00}, gravitational \cite{man97} and statistical physics \cite{cai03}, astrophysics \cite{cra04} and classical mechanics \cite{war04}. Here we show that the Lambert $W$ function can also be used with profit in the study of stochastic  processes. 

This paper is organized as  follows. In Section II, we briefly review the small noise expansion and derive an approximate Fokker--Planck equation in the limit of zero fluctuations. In Sections III and IV,  we  analytically compute  the mean, the variance and the covariance of the random process defined by Eqs.~(\ref{eq1}) and  (\ref{eq01}) in the limit of small noise. We take advantage of the Lambert function to express these quantities in a compact way. Finally, we   use these explicit results to illustrate the complex behavior that arises due to  the interplay of nonequilibrium fluctuations and nonlinearity of the drift. In particular, we show  that the effect of the noise can be greatly enhanced by the nonlinearity. 

\section{Small noise approximation}
\begin{center}
\begin{figure}[h]
\epsfxsize=0.34
\textwidth
\epsffile{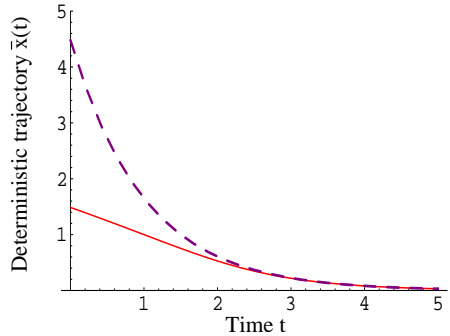}
\caption{Deterministic trajectory $\overline x(t)$, Eq.~(\ref{eq4}), in the absence of noise, with $C=3$ (continuous line). The dashed line represents the exponential approximation valid for large times.}
\label{fig1}
\end{figure}
\end{center}

The small noise approximation is based on the following observation: In the limit $D\rightarrow 0$, where fluctuations are very small, the time evolution of the system is mainly determined by the deterministic trajectory that corresponds to $D=0$. The stochastic trajectory of the system with noise can then be expanded around the deterministic solution. In this way, one obtains an expansion of the Fokker--Planck equation in powers of $\sqrt{D}$ \cite{gar85,kam92}. We begin with the zero noise equation 
\be
 \frac{\pd P(x,t)}{\pd t} =
 -\frac{\pd}{\pd x}\left[K(x)P(x,t)\right] \ ,
\ee
with $K(x)$ given by Eq.~(\ref{eq01}).
The corresponding deterministic equation for the position $x(t)$ of the system can be shown to be 
\be
\label{eq2}
\dot x(t) = K(x) = - \frac{x}{1+x^2}\ .
\ee
An analytical solution of this nonlinear differential equation can be obtained by  introducing the variable $z= x^2$. After integration, we find that $z(t)$ satisfies 
\be
\label{eq3}
z e^z = e^{-2 t+C}\ ,
\ee 
where $C=z_0 +\ln z_0$ is the integration constant. We recognize in Eq.~(\ref{eq3}) the defining equation of the Lambert function and we can therefore write
\be
z(t) = W[e^{-2t+C}]\ .
\ee
As a result, the  solution  of the deterministic differential equation (\ref{eq2}), which we denote in the following by $ \overline x(t) $, is  given by a square root of the $W$ function,
\be
\label{eq4}
\overline x(t) = \pm\sqrt{W[e^{-2t+C}]} \ .
\ee
A typical positive trajectory of this deterministic process is shown in Fig.~(\ref{fig1}). We observe that, as time evolves, the system approaches the stable minimum, $x=0$, of the external potential $U(x) = 1/2\,\ln(1+x^2)$. By using the asymptotic expansions of  the Lambert function for small and large arguments, $W[x]\sim x$ for  $x\ll 1$ and $W[x]\sim \ln x$ for $x\gg 1$, we can  deduce the behavior of the deterministic trajectory $\overline x(t)$ for small and large times.  For small times, we have $\overline x(t) \simeq \sqrt{-2 t+C}$. This nonexponential time dependence corresponds to the nonlinear part of the drift coefficient (\ref{eq01}). On the other hand, for large times, we obtain an exponential decay, $\overline x(t) \simeq \exp[-t+C/2]$, a behavior which is reminiscent of that of an Ornstein--Uhlenbeck process. In this regime,  the system is close  to the origin of the potential  and experiences the linear part of the drift.
\vskip -0.7cm

Having determined the deterministic trajectory in the absence of noise, the next step is now to expand the random trajectory $x(t)$, in the presence of weak  fluctuations,  around the deterministic path $\overline x(t)$. In first order of the small expansion parameter  $ \sqrt{D}$, we write 
\be 
\label{eq5}
x(t) = \overline x(t)+ \sqrt{D}\, y(t) \ .
\ee
Equation (\ref{eq5}) defines a new random variable $y(t)$ with probability density function $Q(y,t)$. By noting  that $Q(y,t)=P(x,t)$, it is straightforward to derive a Fokker--Planck equation for the distribution $Q(y,t)$  from Eq.~(\ref{eq1}). This yields 
\be
\label{eq6}
 \frac{\pd Q(y,t)}{\pd t} =
 -K'(\overline x) \,\frac{\pd }{\pd y}\left[y\, Q(y,t) \right ] +  \frac{\pd^2 Q(y,t)}{\pd y^2} \ ,
\ee
where we have used the first order expansion, $K(\overline x+\sqrt{D}\, y) \simeq K(\overline x) + \sqrt{D}\, y \,K'(\overline x)$. In the limit of small fluctuations, the Fokker--Planck equation (\ref{eq1}) with nonlinear drift is thus transformed into the Fokker--Planck equation (\ref{eq6}) with a {\it linear}, time dependent, drift coefficient, $K'(\overline x(t))\, y$.  
 This is an important point since the lowest cumulants of a time dependent Ornstein--Uhlenbeck process can be calculated in closed form.  
   
\section{Mean and variance}
In this Section, we use the just derived Fokker--Planck equation (\ref{eq6}) with linear drift to evaluate the lowest cumulants of the random variable $x(t)$ within the small noise approximation. The moments $\la x^n\ra$ are defined by the general formula
\be
\la x^n\ra = \int dx\, x^n P(x,t)\ .
\ee
 Of main interest are the first two moments of the distribution $P(x,t)$, from which we can calculate the first two cumulants. The first cumulant, the mean  $\la x\ra$, describes the average displacement of the probability distribution, whereas the second cumulant, the variance $\sigma^2_x= \la x^2\ra -\la x\ra ^2$, is related to the spreading of the distribution due to the noise. It is a simple task to obtain to first two cumulants of $x(t)$ from those of the new variable  $y(t)$ using Eq.~(\ref{eq5}). We find,
\bea
\label{eq7}
\la x\ra &=& \overline x + \sqrt{D}\, \la y \ra \ ,\\
 \sigma^2_x&=& D\,\sigma^2_y\ .
\eea
The mean $\la x\ra$ is hence the sum of the deterministic process $\overline x(t)$ and a small contribution $\sqrt{D}\, \la y \ra$ coming from the noise, while the variance $\sigma^2_x$ is simply proportional to the variance $\sigma^2_y$.  
 
Let us now evaluate the lowest cumulants of the stochastic variable $y(t)$. The time evolution of the cumulants of $y(t)$ is fully determined  by the Fokker--Planck equation (\ref{eq6}). For the first two cumulants, we have \cite{gil92}
\bea
\label{eq8}
\frac{d \la y\ra}{dt} &=& K'(\overline x) \, \la y \ra\ ,\\
\frac{d \sigma_y^2}{dt} &=& 2 K'(\overline x) \,\sigma_y^2 +2 \ .
\eea
A direct  consequence of the linearity of the drift coefficient in Eq.~(\ref{eq6}) is that the above equations are closed for $\la y\ra$ and $\sigma_y^2 $. These first order differential equations  can therefore be  solved,
\bea
\label{eq9}
\la y \ra(t) &=&  \exp \Big[\int_0^t dt_1\, K'(\overline x(t_1))\Big] \ , \\ 
\sigma^2_y(t)&=& 2 \exp\Big[2\int_0^t dt_1\, K'(\overline x(t_1)) \Big]\nonumber \\ 
&\times&\int_0^t dt_1\, \exp\Big[ -2 \int_0^{t_1} dt_2\, K'(\overline x(t_2))\Big] \ .
\eea
We stress that for general nonlinear drift coefficients $K(x)$, these formal integrals cannot be evaluated explicitly.  However, in the present situation, we can use 
the deterministic solution (\ref{eq4}) to obtain the compact expressions
\bea
\label{eq10}
\!\!\la y\ra(t) \!&=&\! \frac{K(\overline x)}{K(x_0)}\ ,  \\
\label{eq10a}
\!\!\sigma^2_y(t) \!&=& \!\frac{2 \overline x^{\,-2} +(x_0^4+12 t -2x_0^{-2})  - \overline x^4}{2}\,K(\overline x)^2 \!\ ,
\eea
where $x_0$ is the initial position of the system. We can easily verify that Eqs.~(\ref{eq10}) and (\ref{eq10a}) reduce to the known results of the Ornstein--Uhlenbeck process when $x_0\rightarrow 0$,
\bea
\label{eq11}
\la y\ra(t)_{OU} &=& \exp[-t]\ ,\\
\label{eq11a}
 \sigma^2_y(t)_{OU}&=& 1-\exp[-2t]\ .
\eea
 In this limit, the system initially starts close to the bottom of the potential and only feels the linear part of drift. The general nonexponential time dependence of Eqs.~(\ref{eq10})--(\ref{eq10a}) thus directly reflects the nonlinear nature of the problem. Moreover, Eqs.~(\ref{eq10})--(\ref{eq10a}) also simplify in the limit of long times, when the system again approaches the origin. Using the asymptotic  properties of the Lambert function, we find for large times
\bea
\la y \ra(t)  &\simeq&  \frac{\exp[-t]}{K(x_0)}\ , \\ 
\sigma^2_y(t)  &\simeq&  1+6 t \exp[-2t] \ .
\eea
Note that in this long time limit, the stochastic process does not exactly reduce to an Ornstein--Uhlenbeck process.

\begin{center}
\begin{figure}[h]
\epsfxsize=0.34
\textwidth
\epsffile{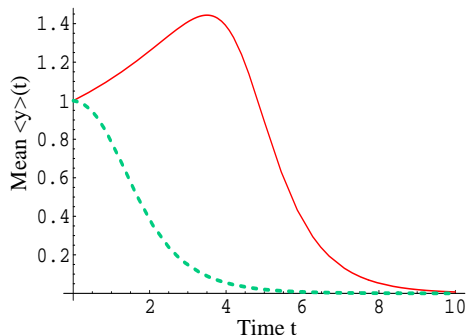}
\caption{Average contribution of the noise $\la y(t) \ra$ as given by Eq.~(\ref{eq10}) for two initial conditions: a) $C = 0.8$ (dotted line) and b) $C = 8$ (continuous line).}
\label{fig2}
\end{figure}
\end{center}
\vspace{-1.3cm}
\begin{center}
\begin{figure}[h]
\epsfxsize=0.34
\textwidth
\epsffile{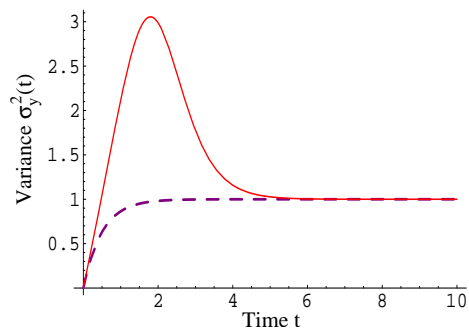}
\caption{Variance $\sigma_y^2(t)$ of the noise contribution, as given by Eq.~(\ref{eq10a}), for $C=3$ (continuous line). The dashed line shows the variance $\sigma^2_y(t)_{OU}$ (\ref{eq11a}) of the Ornstein--Uhlenbeck process.}
\label{fig3}
\end{figure}
\end{center}
\vspace{-0.8cm}

It is  worthwhile 
noticing that $x=1$ is an inflexion point of the potential $U(x)=1/2\,\ln(1+x^2)$: the function $U(x)$ is concave for $x<1$ and convex for $x>1$. This leads to the interesting result that  the influence of the fluctuations on the system  depends on whether the initial position $x_0$ is smaller or larger than one, as illustrated in Fig.~(\ref{fig2}). For $x_0<1$ (or equivalently $C<1$), the time derivative $d\la y\ra /dt$ in Eq.~(\ref{eq8}) is always negative and the average contribution of the noise $\la y\ra(t)$ to the stochastic trajectory in Eq.~(\ref{eq7}) steadily diminishes with time. By contrast, for $x_0>1$ (or $C>1$), the time derivative is first {\it positive} for $t<t_C= (C-1)/2$ and then changes sign. Therefore, the  contribution of the fluctuations to the averaged dynamics first increases with time, reaches a maximum at $t=t_C$ and then decreases  rapidly. We thus have a transient situation where  fluctuations actually push the system away from the deterministic motion. Such an effect does not exist for processes with linear drift. This noise amplification is hence a direct  consequence of the nonlinearity of the drift coefficient.

No such initial condition dependence is observed for the variance $\sigma^2_y(t)$. For all initial positions $x_0$, the variance diplays a similar behavior, as shown in Fig.~(\ref{fig3}): the variance first increases linearly with time and then decays exponentially to one --- the saturation of the variance for large times corresponding to the stationary limit where the effect of the fluctuations is counterbalanced by the confining potential $U(x)$. However, as for the mean $\la y\ra(t)$, a signature of the amplification of the noise by the nonlinearity is clearly visible. We  note indeed that contrary to the Ornstein--Uhlenbeck process, where the variance (\ref{eq11a}) is always smaller than one,  the variance here largely exceeds  the value one. This amplification is found to be stronger the larger the value $x_0$. In other words, the noise amplification is stronger, the farther the system initially starts  from the linear regime. 
\begin{center}
\begin{figure}[h]
\epsfxsize=0.34
\textwidth
\epsffile{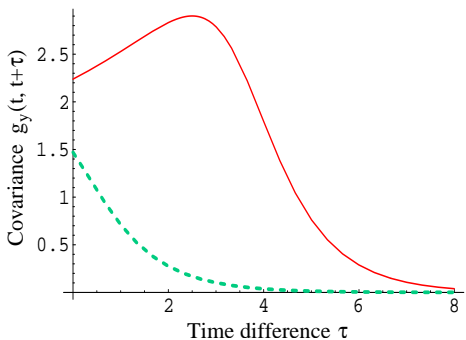}
\caption{Autocorrelation function  $g_y(t,t+\tau)$ of the random variable $y(t)$,  Eq.~(\ref{eq15}), as a function of the time difference $\tau$  for fixed time $t=1$. The continuous line corresponds to an integration constant  $C=8$ and the dashed line to $C=0.8$.}
\label{fig4}
\end{figure}
\end{center}
\vspace{-.7cm}

\section{Autocorrelation function}
We next calculate another quantity of great interest, namely, the autocorrelation function (or covariance), $g_x(t,s)= \la x(t)x(s)\ra-\la x(t)\ra \la x(s)\ra$,  which measures the time  correlations of the process $x(t)$ between different times $t$ and $s$. In the limit $t=s$, the covariance reduces to the variance $g_x(t,t) = \sigma^2_x(t)$. Similar to  the variance, the covariance $g_x(t,s)$ is simply related to the covariance $g_y(t,s)$ through $g_x(t,s)= D\, g_y(t,s)$. The time evolution of the autocorrelation function $ g_y(t,s)$ for the stochastic variable $y(t)$ can easily be determined from the Fokker--Planck equation (\ref{eq6}) and we have \cite{gil92}
\be
\label{eq13}
\frac{d g_y(t,s)}{ds} = K'(\overline x(s))) \,g_y(t,s) \ .
\ee
It is interesting to note that this differential equation for the covariance $g_y(t,s)$ is identical to the differential equation (\ref{eq8}) for the mean $\la y\ra (t)$. This is  an example of a regression theorem for a stochastic process with linear drift, where the one--time and two--time correlation functions obey the same evolution equation \cite{gar85}. The solution of Eq.~(\ref{eq13}) reads
\bea
\label{eq14}
g_y(t,s)&=& \sigma_y^2(t) \exp\left[\int_t^s dt_1\, K'(\overline x(t_1))\right]\nonumber \\
&=&  \sigma_y^2(t)\frac{K(\overline x(s))}{K(\overline x(t))} \ ,
\eea
where the variance $\sigma_y^2(t)$ is given by Eq.~(\ref{eq10a}). In the limit $x_0\rightarrow 0$, Eq.~(\ref{eq14}) reduces to the familiar Ornstein--Uhlenbeck result
\be
g_y(t,s)_{OU} = \left(1-\exp\left[ -2 t\right] \right) \,\exp\left[-\left(s-t\right)\right] \ ,
\ee
as it should. In the limit of long times, the process becomes stationary and the covariance depends only on the time difference $\tau= s-t$,
\be
\label{eq15}
 g_y(t,s) \simeq \exp\left[-\left(s-t\right)\right] \ .
\ee
Equation (\ref{eq15}) indicates an asymptotic exponential  decay  of the time correlations. The behavior of the autocorrelation function for short times is however different,  as shown in Fig.~(\ref{fig4}). In analogy with the mean, two situations have to be distinguished, depending on whether $x_0$ is smaller or larger than one. For $x_0 <1$, the time correlation function $g_y(t,t+\tau)$ always decreases with increasing time difference $\tau $. By contrast, when $x_0>1$, the covariance  first {\it increases} for small $\tau$ before decreasing for larger time differences. We thus have temporary enhancement of the temporal correlations of the random process. This is another unexpected consequence due to the nonlinearity of drift term.  
\vspace{-.3cm}

\section {Conclusion}
In summary, we have obtained explicit results for the mean, the variance and the covariance of a stochastic process described by a Fokker--Planck equation with nonlinear drift in the small noise approximation. We were able to write down compact analytical expressions for these physical quantities by using the Lambert $W$ function. This is hence an example of a random  process whose study is notably facilitated by the use of the Lambert function. Moreover, we have  found that, far from equilibrium, the nonlinearity of the drift leads both to a transient amplification of the noise and  to a transient increase of the temporal correlations. These effects illustrate the rich behavior that can result from the interplay of noise and nonlinearity. 

We acknowledge financial support from the Landesstiftung Baden--W\"urttemberg.  


\begin{thebibliography}{99}
\bibitem{rem} We only consider the case of constant diffusion coefficient, since any Fokker--Planck equation can be reduced to the form (1), see Ref.~[1].
\end{thebibliography}
\end{document}